\documentstyle[12pt]{article}
\def\uglu{\hskip 0pt plus 1fil minus 1fil}
\def\uglux{\hskip 0pt plus .75fil minus .75fil}

\def\slashed#1{\setbox200=\hbox{$ #1 $}
    \hbox{\box200 \hskip -\wd200 \hbox to \wd200 {\uglu $/$ \uglux}}}

\def\fA{{\cal A}}
\def\fE{{\cal E}}
\def\fH{{\cal H}}
\def\fL{{\cal L}}

\newcommand{\lsim}{\mathrel{\lower4pt\hbox{$\sim$}}
\hskip-12.5pt\raise1.6pt\hbox{$<$}\;}

\newcommand{\gsim}{\mathrel{\lower4pt\hbox{$\sim$}}
\hskip-12.5pt\raise1.6pt\hbox{$>$}\;}

\begin{document}

\setlength{\textheight}{7.5truein}

\def\vtd{$V_{td}$}
\def\what{\underbar{\hskip.5in}}

\def\lsim{\mathrel{\lower4pt\hbox{$\sim$}}
\hskip-12pt\raise1.6pt\hbox{$<$}\;}

\def\gsim{\mathrel{\lower4pt\hbox{$\sim$}}
\hskip-12pt\raise1.6pt\hbox{$>$}\;}

\begin{titlepage}
\noindent\hspace*{10cm}SLAC-PUB-95-6877\\
\noindent \hspace*{10cm}BNL-AS-5495\\

\vfill
\begin{center}

{\bf Neutral Higgs CP Violation at $\mu^+\mu^-$ colliders}\\

\vfill
{David Atwood$^a$ and  Amarjit Soni$^b$}\\
\end{center}
\vfill

\begin{flushleft}

a) Stanford Linear Accelerator Center,
Stanford University,
Stanford, CA\ \ 94309, USA \\
b) Department of Physics, Brookhaven National Laboratory,
Upton, NY\ \ 11973, USA\\

\end{flushleft}

\vfill

\begin{quote}
{\bf Abstract}:
CP violating asymmetries in both the production and
the decays of
$s$-channel neutral Higgs bosons at $\mu^+\mu^-$ colliders
can be large.
If the CP violation occurs in the muon-Higgs coupling,
one  observes a
production asymmetry with  transversely polarized $\mu^+\mu^-$ beams.
Likewise if the CP violation occurs in the top-Higgs  or $\tau$-Higgs
coupling, one  observes an azimuthal decay asymmetry in  $\mu^+\mu-\to
t\bar t$ (or $\tau^+\tau^-$). CP studies at such colliders allow
a uniquely clean way of deducing the underlying CP phases; in
turn these parameters would significantly improve our understanding
of baryogenesis.
\end{quote}

\vfill

\begin{center}

Submitted to {\it Physics Letters B}

\end{center}

\vfill

\hrule
\vspace{5 pt}
\noindent
* This work was supported by US Department of Energy contracts
DE-AC03-765F00515 (SLAC)  and DE-AC02-76CH0016 (BNL).

\end{titlepage}

The Large Hadron Collider (LHC) currently being planed at  CERN may
provide a reach for new physics of up to a few TeV\null. It is to be
hoped that this hadronic capability will be complemented with
leptonic machines.

Due to synchrotron radiation effects it is  probably impractical to
construct an $e^+ e^-$ storage ring  (such as LEP) at these higher
energies.  Two possible leptonic colliders at the TeV energy scale that
have been suggested are  a colliding beam $e^+ e^-$ machine (Next Linear
Collider or NLC),   which would operate on principles similar to the
current SLC   and a Muon Collider (MUC).  Since synchrotron radiation is
so much smaller at a MUC compared to an electron machine, a MUC has the
advantage that muon beams could be re-circulated in a storage ring and
thus  muons could be  reused several hundred times before their eventual
decay.  Consequently MUCs have been receiving increasing attention in
the past few years \cite{cline}--\cite{barger}.  By the same token, NLC
has been the subject of intense research effort \cite{epemcol} in the
past decade and its feasibility is no longer widely questioned whereas
for MUCs  many challenges in accelerator design still need to be
overcome \cite{cline}--\cite{palmer}.

For most physics applications muon collisions and electron collisions
are virtually identical. Hence factors such as cost, and the
engineering problems involved in building such colliders are likely to
be the determining factor in their eventual construction. If a MUC
is eventually constructed, a possible physics bonus may be the
ability to study neutral Higgs bosons in the $s$-channel
\cite{cline,barger}.

In this paper, we shall assume that at least one neutral Higgs scalar,
which we will denote $\fH$, will be discovered in the mass range
$100\leq m_\fH\leq 1000 GeV$ and that the muon collider will be tuned to
study such a resonance in the $s$-channel so that $\sqrt{s}=m_\fH$. In
this way  the feeble coupling of the $\mu$ to the Higgs may be
partially compensated by the  resonance enhancement and so the signal of
an $s$-channel Higgs boson may be studied in detail. Bearing this in
mind we consider how the CP phase of its couplings to fermions may be
investigated. We find that the method gives a uniquely clean way of
deducing CP violating phases in the coupling of the neutral Higgs to
leptons and to the top quark.

As is well known, at the present time, the only form of CP violation
experimentally observed is in the $K^0$ system. It is thought that these
results can be explained through the standard model's phase in the CKM
matrix \cite{kobayshi},  a hypothesis which will be tested at $B$
factories now being built.  However, strong theoretical arguments exist
that suggest that additional forms of CP  violation are needed to
explain the baryon asymmetry of the universe \cite{cohen}. Indeed one
class of models which has been  shown to give reasonable baryon
asymmetries are models involving CP violation in the Higgs sector.

Motivated by these considerations we shall investigate
two methods of
observing a CP violating signal in such a Higgs resonance
at a MUC\null. First, if the $\mu$ beams are transversely polarized at
different inclinations, there is a CP violating production asymmetry
proportional to $\sin\phi_\mu$, $\phi_\mu$  being the azimuthal angle
between the polarization directions. This is sensitive to CP violation
in the  Higgs-$\mu\mu$ coupling.  The second observable we consider is
an analogous asymmetry in   $\mu^+\mu^-\to t\bar t$ (or $\tau^+\tau^-
$).  In the latter case, for instance, the signal  is proportional to
$\sin \phi_t$ where $\phi_t$ is the azimuthal angle between  the
polarization of the top quarks; $\phi_t$ may easily be determined if the
tops decay leptonically. Polarization of the colliding muons is not a
necessity for  this study, although longitudinal polarization of the
beam can be used to  reduce the  background. In this case one is
sensitive to CP violation in the Higgs-$t\bar t$ (or $\tau^+\tau^-$)
coupling which relates directly to some proposed mechanisms of
baryogenesis in the early universe \cite{cohen}.

In all three cases mentioned in the preceding
paragraph the asymmetries
may be large. However, to a certain extent, they will be obscured by
standard model backgrounds, as we will discuss below. We find that
observation of these asymmetries will certainly be possible if  the
Higgs is relatively light or narrow leading to a relatively large
production rate. Furthermore, in many theoretical scenarios, even if the
Higgs is not that light, it can have a large production rate and it can
lead to appreciable asymmetries.

Although the underlying model we have in mind may contain several  Higgs
doublets we would like to cast our analysis in a relatively model
independent way. Let us assume that a single Higgs is under experimental
study which we will denote $\fH$. We will parameterize its coupling to a
fermion $f$ by

\begin{equation}
C_{\fH ff}=C^0_{ff}\chi_f e^{i\gamma_5 \lambda_f}
\end{equation}

\noindent where $C^0_{ff}$ is the coupling in the standard model (with
one Higgs doublet). In such theories with an extended Higgs sector the
field $\fH$ may either be a scalar $H$ or a pseudoscalar $A$. If $\fH=A$
then it does not couple to  two gauge bosons while if $\fH=H$ we will
parameterize the  coupling to two vector bosons  ($VV=ZZ$ or $WW$) as:

\begin{equation}
C_{\fH VV}=C^0_{VV} \cos\alpha .   \label{two}
\end{equation}

\noindent where $C^0_{VV}$ is the coupling of the SM Higgs to the gauge
bosons and $\alpha$ is the angle between the observed
Higgs $\fH$ and
the orientation of the vacuum in the Higgs space.

To serve as an illustration we will consider three sample models for the
$\fH$ in question:

\begin{itemize}
\item{1:}  $\fH=H$ with  $\alpha=\lambda_f=\pi/4$ and $\chi_f=1$ for all
fermions.

\item{2:}  $\fH=A$ with $\lambda_f=\pi/4$ and $\chi_f=1$ for all
fermions.

\item{3:}  $\fH=A$ with $\lambda_f=\pi/4$ and $\chi_l=\chi_d=5$  and
$\chi_u=1/5$.

\end{itemize}

\noindent The last case is motivated by models of the  type that, for
instance, are used for the Higgs sector for supersymmetric (SUSY)
scenarios \cite{barger}.  In such a case one might want to partially
explain the large top mass  by making the corresponding vacuum
expectation value (vev) large giving rise to
$\chi_l=\chi_d=v_2/v_1=1/\chi_u$ for large $v_2/v_1$. Although minimal
SUSY models do not have CP violation in the  Higgs sector, more
complicated Higgs sectors or  non-minimal SUSY models do readily admit
CP violation \cite{grzad}.

It is  difficult to observe the effects of the Higgs resonance except if
the accelerator is precisely on shell \cite{cline,barger}. If $s$ is the
center of mass energy and $s=m_\fH^2$ then  defining $B_\mu={\rm
Br}(\fH\to\mu^+\mu^-)$ and  $\sigma_\fH=\sigma(\mu^+\mu^-\to\fH)$, the
cross section for Higgs production is:

\begin{equation}
\sigma_\fH = {4\pi\over m_\fH^2}B_\mu.
\end{equation}

\noindent It is useful to compare this with $\sigma_0 =\sigma(\mu^+
\mu^-\to \gamma^* \to e^+e^-)$  so we define

\begin{equation}
R(\fH) = {\sigma_\fH \over \sigma_0} = {3\over\alpha_e^2} B_\mu
\label{rdef}
\end{equation}

\noindent where $\alpha_e$ is the electromagnetic coupling. It is clear
in the above that  the key to having a large rate for Higgs production
is for its width to  be as small as possible (thereby increasing
$B_\mu$). The main possible decay modes of the Higgs which we consider
are $\fH\to ZZ$, $\fH\to W^+W^-$, $\fH\to t\bar t$, $\fH\to b\bar b$,
and $\fH\to\tau^+\tau^-$. The last two become important for  the case
when $m_\fH < 2 m_W$. The decay rates to these modes, given the above
couplings, can be readily calculated by using the results that exist in
the  literature \cite{gunion}:

\begin{eqnarray}
\Gamma(\fH\to t\bar t) &=& {3 g_W^2 m_t^2 m_\fH
\over 32\pi m_W^2}
\beta_t\left[\beta_t^2+(1-\beta_t^2)\sin\lambda_t\right] \chi_t^2 K_t
\nonumber \\
\Gamma(\fH\to b\bar b) &=&  {3 g_W^2 m_b^2 m_\fH
\over 32\pi m_W^2} \chi_b^2
K_b\nonumber \\
\Gamma(\fH\to ZZ) &=& {g^2\over 128\pi}{m_\fH^3\over m_Z^2} \beta_Z
(\beta_Z^2+12{m_Z^4\over m_\fH^4})\cos^2\alpha \nonumber \\
\Gamma(\fH\to WW) &=& {g^2\over 64\pi}{m_\fH^3\over m_W^2} \beta_W
(\beta_W^2+12{m_W^4\over m_\fH^4})\cos^2\alpha
\end{eqnarray}

\noindent where $\beta_i=\sqrt{1-4m_i^2/m_\fH}$. The $ZZ$ and $WW$ modes
are only relevant if $\fH=H$. In these equations $K_b$, $K_t$ are QCD
corrections given, for example, in Ref.~\cite{gunion}.

In comparison, the width of the Higgs decay to $\mu^+\mu^-$ is

\begin{equation}
\Gamma(\fH\to \mu^+\mu^-) = \frac{g_W^2 m_\mu^2 m_\fH}
{32\pi m_W^2}\chi_\mu^2
\end{equation}

\noindent In Figure~1 we present the value of $R(\fH)$ at  $s=m_\fH^2$
assuming that $\alpha=45^\circ$ for the  $\fH=H$ and $\fH=A$  cases  as
a function of $m_\fH$.

In deriving equation (\ref{rdef}) above we assumed that the energy of
the collider could be controlled to a precision much finer than the
width of the Higgs. While this should be generally true  if $m_\fH \gg
2m_W$ or $2m_t$,  below this threshold it may well not be the case. To
take this into account let us assume that the actual value of $s$ is
uniformly distributed in the range:

\begin{equation}
m_\fH^2(1-\delta)<s<m_\fH^2(1+\delta)
\end{equation}

\noindent The beam spread thus defined will lead to an observed rate of
Higgs production described by:

\begin{equation}
\tilde R(\fH)=\left[ \frac{\Gamma_\fH}{m_\fH\delta} \arctan
\frac{m_\fH\delta}{\Gamma_\fH} \right] R(\fH) \label{rtildef}
\end{equation}

\noindent In Figure~1 we also show $\tilde R(\fH)$ using the value of
$\delta=10^{-3}$.

Let us now consider the production of $\mu^+\mu^-\to\fH$  with
transversely polarized $\mu$ beams. We take the $z$-axis  in the center
of mass frame  to be the direction of the  $\mu^-$ beam to be polarized
in the $x$ direction. Assume further that the polarization of the
$\mu^+$ beam is inclined at an angle of $\phi_\mu$. The cross section
in this case becomes:

\begin{equation}
\sigma(\phi_\mu)=(1 -\cos 2\lambda_\mu \cos\phi_\mu +\sin 2\lambda_\mu
\sin\phi_\mu )\sigma_0
\end{equation}

\noindent where $\sigma_0$ is the unpolarized cross section. Then the CP
odd asymmetry in the production will be given by:

\begin{equation}
A_\mu\equiv \frac{\sigma(90^\circ)-\sigma(-90^\circ)}{\sigma(90^\circ)
+\sigma(-90^\circ) } =\sin 2\lambda_\mu
\end{equation}

\noindent Thus the underlying CP violating phase $\lambda_\mu$ becomes
cleanly measurable. Since this phase ($\lambda_\mu$)is unconstrained,
large asymmetries approaching 100\% are possible. As we shall discuss
below, this asymmetry will be diluted somewhat by the standard
model backgrounds.

Let us now turn our attention to the analogous type of asymmetry in the
decays $\fH\to t\bar t$ or $\tau^+\tau^-$ \cite{cpviol}. In order to
observe a  CP violating correlation in the spins of the top quarks or
$\tau$-leptons, we clearly need to look at the  correlations among their
decay products.

Consider first the determination of the spin of a fermion in general.
Suppose $f$ decays to $XY$ where $X$, $Y$  may be a single particle or a
multiparticle state. Let $\epsilon_X^f$ be the ``analyzing power'',
i.e., the degree to which the momentum of $X$ is correlated with the
spin of  $f$, defined by

\begin{equation}
\epsilon^f_X\equiv 3<cos\theta_X>.
\end{equation}

\noindent where $\theta_X$ is the angle between the spin of $f$  and
$\vec P_X$, in the $f$ rest frame.

We discuss the decay of the top quark first.  If the top decays
semi-leptonically, $t\to bW\to b{l^+\nu_l}$  (for $l=e$, $\mu$) then
$\epsilon^t_l=1$ \cite{atwood}. For hadronic decays of the top one could
obtain $\epsilon=1$ if one could tag which jet was the $\bar d$ type
(thus playing the role of the positron). Experimentally this is hard to
do. We will therefore only use the inclusive $W$ momentum to determine
the polarization for the hadronic decays. The  analyzing power in this
case is thus:

\begin{equation}
\epsilon^t_W= {m_t^2-2m_W^2 \over m_t^2+2m_W^2}  \approx 0.39
\label{epstw}
\end{equation}

Next consider the case of the $\tau$ lepton. If the $\tau$ decays
semi-leptonically, then  $\epsilon^\tau_l=-\frac{1}{3}$ \cite{braaten}
(note that if one could observe the $\bar\nu_l$ then $\epsilon=1$).  In
\cite{braaten}  it is shown that the inclusive hadronic decays of the
$\tau$ can give $\epsilon^\tau_h=-.42$ ($h$ stands for inclusive hadron
states).  If one analyses the detailed structure of the hadronic decays
one can obtain  better results \cite{tsai,kuhnone}
but we will
just use this number in our discussions to follow.

Let us define a coordinate system in the Higgs center of mass frame
where the $z^\prime$ axis is in the direction of the $f$ (i.e., $t$ or
$\tau$) momentum. Let us now consider the $f$ decays via $f\to X_iY_i$
and the $\bar f$ decays $\bar f\to \bar X_j \bar Y_j$. We define the
angle $\phi_{ij}$ to be the azimuthal angle  between the $p_{Xi}$ and
the $p_{\bar Xj}$ projected into the  $x^\prime-y^\prime$ plane:

\begin{equation}
\sin(\phi_{ij}) = \frac{\vec p_{Xi}\times \vec p_{\bar Xj}\cdot \vec
p_f}{|\vec p_{Xi}|\ | \vec p_{\bar Xj}|\ |\vec p_f|}
\end{equation}

Taking into account $\epsilon^f_i$, $\epsilon^f_j$ the differential
distribution  in $\phi_{ij}$ is (note that $\epsilon^{\bar
f}=-\epsilon^f$):

\begin{equation}
\frac{d\Gamma}{\Gamma\ d\phi_{ij}} = 1 + \frac{\pi^2}{16}
\epsilon^f_i\epsilon^f_j \rho_f \cos 2\lambda_f \cos\phi_{ij} +
\frac{\pi^2}{16} \epsilon^f_i\epsilon^f_j \eta_f \sin 2\lambda_f
\sin\phi_{ij} \label{etadef}
\end{equation}

\noindent  In eqn.~(\ref{etadef}), for $f=t$ (so that the threshold mass
effects must be  retained), $\rho$ and $\eta$ are given by:
\begin{eqnarray}
\rho_t &=&
{1-\beta_t^2-(1+\beta_t^2)\cos 2\lambda_t
\over
\cos 2\lambda_t(1+\beta_t^2-(1-\beta_t^2)\cos 2\lambda_t)}
\nonumber \\
\eta_t &=& \frac{\beta_t}{1-(1-\beta_t^2)\cos^2\lambda_t} \label{ptetat}
\end{eqnarray}
\noindent On the other hand, if $f=\tau$ then the above simplifies to
$\rho_\tau=\eta_\tau=1$. The resultant CP violating asymmetry which we
will consider is:

\begin{equation}
A^f_{ij} = \frac{\Gamma(\sin\phi_{ij}>0)-\Gamma(\sin\phi_{ij}<
0)}{\Gamma (\sin\phi_{ij}>0)+\Gamma(\sin\phi_{ij}<0)}  \label{afij}
\end{equation}

\noindent Thus using the above distribution we find:

\begin{equation}
A^f_{ij}=\frac{\pi}{8}\epsilon^f_i\epsilon^f_j\eta_f\sin 2\lambda_f
\end{equation}

Of course we would like to optimally combine the resultant asymmetries
from different pairs of modes $\{ij\}$. In order to do this we would
like to choose weights $w_{ij}$  normalized by:

\begin{equation}
\sum w_{ij}B_iB_j=\sum B_iB_j
\end{equation}

\noindent
where $B_i$ is the branching ratio of $f\to i$
and the summation is over all the modes under observation.
Thus
the total asymmetry

\begin{equation}
\fA^f=\sum w_{ij}A^f_{ij} B_iB_j
\end{equation}

\noindent is
to be
maximized. This is done by taking $w_{ij}\propto A_{ij}^f$
\cite{opt}.
Applying this to the asymmetries defined in (\ref{afij}) we find that
the total asymmetry is given by

\begin{equation}
\fA^f=\frac{\pi}{8}(\fE^f)^2\eta_f\sin 2\lambda_f \label{afijdef}
\end{equation}

\noindent where

\begin{equation}
\fE^f=\sqrt{ \sum B_i (\epsilon^f_i)^2 }
\end{equation}

\noindent Selecting the top quark and the $\tau$ lepton modes mentioned
above we obtain numerically (using $m_t=176GeV$, $B_l^\tau=.36$,
$B_h^\tau=.64$,  $B^t_e+B^t_\mu={2\over 9}$ and $B^t_h+B^t_\tau={7\over
9}$):

\begin{equation}
\fE^t=.58 \qquad \fE^\tau=.39
\end{equation}

In the case of top decay, if the quark or the anti-quark decay
semi-leptonically, the 4-momenta of the  undetected  neutrino(s) may be
reconstructed by using the mass constraints   of the top and the $W$.
For the decay of the $\tau$ lepton, it is less obvious that this can be
done. For instance, if one or both of the $\tau$ leptons decays
leptonically, there are at least 3 $\nu$'s in the final state and the
kinematics does not provide sufficient restrictions to reconstruct the
event. Even if both $\tau$'s decay hadronically, although it would
appear that there is enough information to reconstruct the event, in
fact solving the kinematics gives rise to a ``quadratic ambiguity''
resulting in  an ambiguity in the sign of $\sin \phi_{ij}$.  Therefore
with this information alone the asymmetry $\fA^f$
cannot be evaluated.

These problems with the $\tau$ may be resolved as follows given a
reasonable vertex detection capability.

Suppose $X_i$ is a particle emerging from the $\tau^-$ and $\bar X_j$ is
a particle emerging from the $\tau^+$. In general the tracks of these
two particles will not intersect. Let us define the vector $v_{ij}$ to
be the displacement from the track of $X_j$ to the track of $X_i$ at
either their closest approach or the   closest approach inferred from
extrapolating the portion of the tracks which may be observed. Typically
the magnitude of $v_{ij}$ will be of the order of $2c\tau_\tau=180\mu m$
(in spite of the fact that the displacement of the decaying $\tau$'s is
of order  $2 c \tau_\tau \gamma= O(1cm)$). We then define the angle
$\tilde\phi_{ij}$ to be the  difference in the azimuthal angle between
$\vec p_{Xi}$ and  $\vec p_{\bar Xj}$  with respect to an axis in the
$v_{ij}$ direction.

Recall that the asymmetry $A_{ij}$ is just the expectation value of
$\sigma(\sin(\phi_{ij}))$ where:

\begin{equation}
\sigma(x)=\left\{
\begin{array}{lcr}
+1&{\rm if}&x>0\\
-1&{\rm if}&x<0
\end{array}
\right.
\end{equation}

\noindent A vertex detector may allow the determination of
$\sigma(\sin\phi)$ in  a $\tau^+\tau^-$ event using one or more of the
following four methods:

\begin{itemize}
\item{1:} Since the displacement of the $\tau$ decay from the
interaction point is on the order of  $c \tau_\tau \gamma= O(1cm)$
it is not unreasonable to assume that, depending on detector
design, in many events either the  $\tau^+$ or the $\tau^-$  will form a
track in the vertex detector. If this is the case then the $z^\prime$
axis is determined and $\phi_{ij}$ is obtained directly from the momenta
of $X_i$ and $\bar X_j$.

\item{2:} If $X_i$ or $\bar X_j$ consist of multiple charged tracks,
running these tracks back to their common decay vertex again determines
the  $z^\prime $ axis if the machine is configured so that the
interaction region is in a single point.
More generally
if the interaction region is
not a well defined point,
one may use the decay vertex and the combined
momentum of the multiple tracks to define the trajectory of the center
of gravity of the system and use method (3) below treating this
trajectory as an effective single track.

\item{3:} If both $X_i$ and $\bar X_j$ are single charged tracks (e.g.,
$\tau^-\to \mu^-\bar \nu_\mu\nu_\tau$ or $\tau^-\to \pi^-\nu_\tau$) then

\begin{equation}
\sigma(\sin\phi_{ij})=\sigma(\sin\tilde\phi_{ij}) \label{ident}
\end{equation}

\noindent so that the exact decay vertex need not be identified. To see
why (\ref{ident}) should be true, let $\vec x_1$ be any point on  the track
of $X_i$ and $\vec x_2$ be any point on the track of  $\bar X_j$. If $\vec
u=\vec{x_1}-\vec{x_2}$ then  $\vec u \cdot (\vec p_i \times \vec p_j) =\vec
v_{ij}
\cdot (\vec p_i \times \vec p_j)$. If the $\tau^-$ decays at time  $t_1$
and the $\tau^+$ at time $t_2$ then the vertex points $\vec V_1$,
$\vec V_2$ are
$\vec V_1=m_\tau^{-1}t_1\vec p_{\tau^-}$  and $\vec V_2=m_\tau^{-1}t_2\vec
p_{\tau^+}$ therefore
$\vec V_1-\vec V_2=m_\tau^{-1}(t_1+t_2)\vec p_{\tau^-}$ for
the $\tau^-$ and $\tau^+$ respectively. Since the point $\vec V_1$ lies on
the $X_i$ track  and the point $\vec V_2$ lies on the $X_j$ track,

\begin{eqnarray}
\sigma(\sin\tilde\phi_{ij}) &=& \sigma(\vec v_{ij} \cdot \vec p_i \times
\vec p_j)\nonumber \\
&=& \sigma( (\vec V_1-\vec V_2) \cdot \vec p_i \times \vec p_j)\nonumber \\
&=& \sigma(\vec  p_\tau^- \cdot \vec p_i \times \vec p_j)\nonumber \\
&=& \sigma(\sin\phi_{ij})
\end{eqnarray}

\item{4:} If both $\tau$'s decay hadronically
we can use the method of \cite{kuhntwo}.
In this instance
the hadronic states will
contain charged tracks
$X_i\to X_{i,1}\dots X_{i,n}$ and $\bar
X_j\to X_{j,1}\dots X_{j,m}$. For any pair of charged tracks
$\{X_{i,k},\bar X_{j,l}\}$ we define  $\tilde\phi_{i,k\ j,l}$  as above.
In this case we first kinematically solve for the neutrino momenta up to
the ``quadratic ambiguity'' mentioned above. Note that this is in fact a
parity ambiguity so that only one of the reconstructions will give the
correct values for   $\sin \tilde\phi_{i,k\ j,l}$.  The false
reconstruction will give exactly the wrong sign for each of these angles
providing  multiple  checks on the parity of the reconstruction.

\end{itemize}

The only decay combination of modes which is not obviously susceptible
to methods 2--4
(although method 1 will work in all cases given a
vertex detector sufficiently close to the beam axis)
is $X_i=\pi^-\pi^0$ and $\bar
X_j=l^+$ (or vice versa). Let us assume  that the $2\pi$ mode is
dominated by the $\rho$ resonance. Since, using equation (\ref{epstw})
generalized to  $\tau\to\rho\nu$ gives $\epsilon^\tau_\rho=.45$ so that
$\epsilon^\tau_\rho \epsilon^\tau_l \approx.15\approx(\fE^\tau)^2$. Thus
totally ignoring these modes is roughly equivalent to  reducing the
cross section by  $2 B^\tau_\rho B^\tau_l\approx 10\%$.

The above asymmetries must be observed against the CP even standard
model backgrounds $\mu^+\mu^-\to b\bar b$, $t\bar t$, $ZZ$ and $W^+W^-$.
Therefore, using the expressions for the rates for these reactions given
in Ref.~\cite{bargertwo}, we numerically study their impact on the CP
violating signals that we are seeking.

When looking for decay asymmetries of the Higgs the ability to control
the longitudinal polarization  of the beams may help to increase the
signal over the background \cite{simsug}. This is because the Higgs
production flips the chirality of the muon while all standard model
interactions preserve it. Thus, for instance, if both beams are left
polarized with polarization $P$ then the Higgs production is enhanced by
$1+P^2$ while the background is reduced by $1-P^2$ \cite{simsug}. In the
numerical results below we will illustrate the effects of polarization
for a few cases.

Let $R^0_j$ be the standard model contribution to a specific final state
$j$. Let  $\fL_0$ be the integrated luminosity per year of a given
accelerator. For a final state $f$  let us denote by $y_f^{(3\sigma)}$
the number of years needed to accumulate a  3-$\sigma$ signal for the CP
asymmetry. This will be given by

\begin{equation}
y_f^{(3\sigma)} = 9\frac{ R^0_j +R_\fH  Br(\fH\to j) }{A_j^2 R_\fH^2
Br(\fH\to j)^2 \sigma(\mu^+\mu^-\to e^+e^-)\fL_0}
\end{equation}

\noindent We will use the notation  $y^{3\sigma}_i$  for the CP
asymmetry in the production  obtained by monitoring the decay of the
Higgs to the final state $i$ and  $\hat y^{3\sigma}_j$  for the
azimuthal asymmetries described above by observing the decay of the
state $j$ (i.e., $j=t\bar t$ or $\tau\tau$).

In the following we will take the luminosity to be  $10^{34}\ {\rm
cm}^{-2} {\rm s}^{-1}$ so assuming that a year is  $10^7 {\rm s}$
(taking ${1/3}$ efficiency),  $\fL_0=10^{41}\ {\rm cm}^{-2} yr^{-1}$.
Numerical results for the three cases of  Higgs couplings given below
eqn.~(\ref{two}) are summarized in Figures 1--4. For definiteness in all
the cases we will take $\lambda_f=\pi/4$.

In Figure~1 we consider the overall Higgs production cross sections in
relation to $\mu^+\mu^-\to\gamma^*\to e^+e^-$. In each of the three
cases we show $R_\fH$ and $\tilde R_\fH$ with $\delta=10^{-3}$. Note the
pronounced effect of the vev ratio in case (3) due to the fact that the
decay to $t\bar t$ is suppressed and the $b\bar b$ and  $\mu\mu$ modes
are enhanced.

In Figure~2 we consider $y^{3\sigma}$ for production asymmetries as
monitored by various final states in case (1). Below the $WW$ threshold
$y^{3\sigma}$ is in the $10^{-3}-10^{-4}$ range so observation should
be relatively easy. Indeed, in this instance, even a machine luminosity
of $10^{31}$--$10^{32}$ cm$^{-2}$s$^{-1}$ may be adequate. Above this
threshold  $y^{3\sigma}_{WW}$ rapidly increases and passes through $1$
at  $m_\fH\approx 250$ GeV\null.

In Figure~3 we consider the same quantities for case (2). Here we take
$\delta=10^{-3}$ in all the curves.  In this case the $WW$ and $ZZ$
thresholds are absent so a small value of  $y^{3\sigma}$  persists up to
the $t\bar t$ threshold. So again for this mass range the production
asymmetry is a very sensitive probe and observation may be feasible even
with a machine of $\fL\sim10^{32}$ cm$^{-2}$s$^{-1}$. After the $t\bar
t$ threshold   $y^{3\sigma}_{t\bar t}$ rapidly grows  passing through
$1$ around  $m_\fH\approx 400$ and observation starts to become rather
difficult. We also show the decay asymmetries  $\hat y^{3 \sigma}_{\tau
\tau}$  and  $\hat y^{3\sigma}_{t\bar t}$.  $\hat y^{3\sigma}_{\tau
\tau}$  is about 10 below the $t\bar t$ threshold.   $\hat y^{3
\sigma}_{t\bar t}$ grows rapidly from 1 to 100 above threshold. The
effect of polarization is also illustrated in this figure for these two
instances of $\hat y^{3\sigma}_{\tau \tau}$ and $\hat y^{3\sigma}_{t
\bar t}$. The lower dotted and dashed dotted curves are with $P=0.9$
(the upper ones with $P=0$). We thus see that polarization can improve
the effectiveness of the decay asymmetries by about one order of
magnitude.

In Figure~4 we plot the same quantities for case (3). In this case
$y^{3\sigma}$ ranges between $10^{-5}$ and $10^{-3}$ over the entire
mass range so the production asymmetries are excellent probes for
$m_{\fH} \lsim800$ GeV\null. Indeed for this case the production
asymmetries appear detectable even with $\fL\sim10^{31}$
cm$^{-2}$s$^{-1}$. The decay asymmetry  $\hat y^{3\sigma}_{t\bar t}$ is
around $10^{-2}$ and  $\hat y^{3\sigma}_{\tau\tau}$  is in the range
.1--1 below the $t\bar t$ threshold  and 1--100 above it.
We have also shown
the effect of using polarization $P=.9$ on
$\hat y^{(3\sigma)}_{\tau\tau}$ and $\hat y^{(3\sigma)}_{tt}$. Again
polarization seems to improve the situation by about an order of
magnitude. Thus the decay asymmetry in the $\tau\tau$ channel is quite
effective for $m_{\fH}\lsim 300$ GeV; for a heavier $\fH$ the $t\bar t$
channel seems very promising and $\fL\sim10^{32}$ cm$^{-2}$s$^{-1}$ may
well be adequate.

Summarizing, this study shows that production asymmetries represent a
promising way to  study CP violation in the couplings of $\fH$ in all
the  three cases for at least some  range of Higgs masses.
Decay
asymmetries appear less effective and probably are most useful in models
where there is a large  ratio between the vacuum expectation values.
Longitudinal polarization of the muon beams helps to significantly
reduce the backgrounds to the decay asymmetries and can render them
quite viable in a variety of Higgs models.

\bigskip

This work was supported by US Department of Energy contracts
DE-AC03-765F00515 (SLAC)  and DE-AC02-76CH0016 (BNL).

\pagebreak
\bigskip
\noindent{\bf Figure Captions}
\medskip

\bigskip
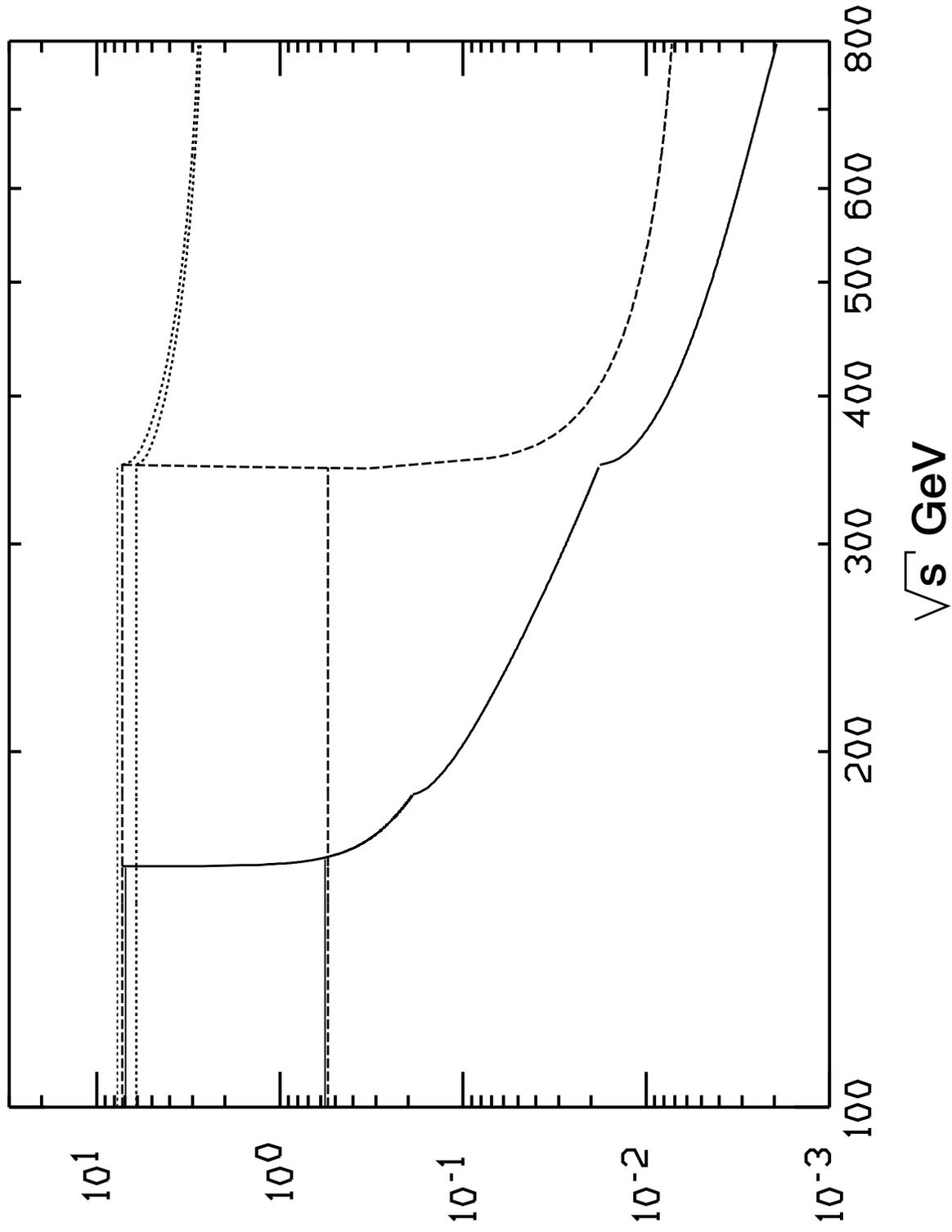
\noindent{Figure 1:}
\medskip

The value for $R_\fH$ as a function of $m_\fH$ for the following three
cases: (1) $\fH=H$,  $\alpha=\pi/4$ and $\chi_f=1$ (solid) (2) $\fH=A$
and $\chi_f=1$ (dashes) (3) $\fH=A$ and $\chi_l=\chi_d=5$ and
$\chi_u=1/5$  (dots). In each case the upper branch represents  the
result for $\sqrt{s}=m_\fH$ while the lower branch  is the result with
an energy spread given by $\delta=10^{-3}$.

\bigskip
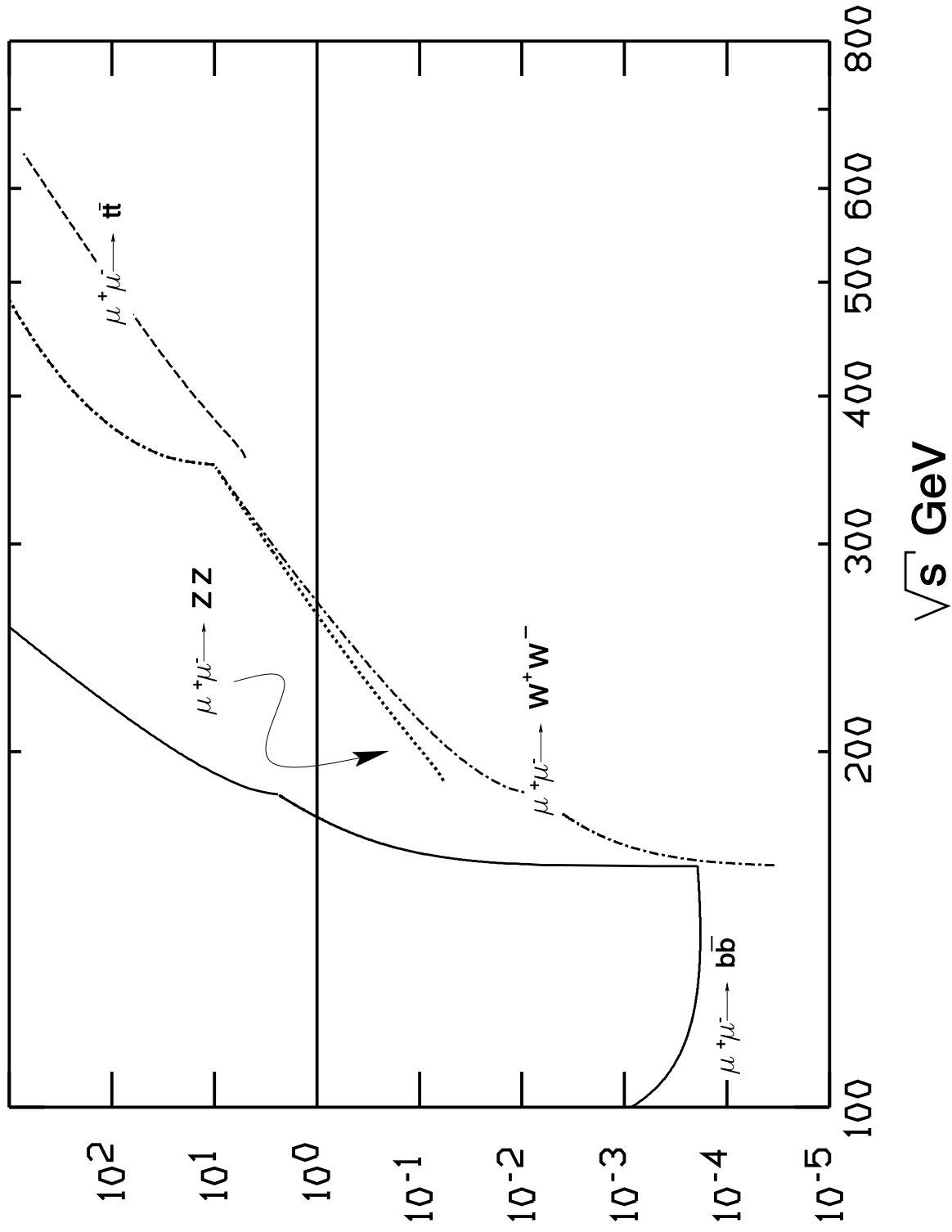
\noindent{Figure 2:}
\medskip

The value of $y^{(3\sigma)}$ for case 1 (see caption to Figure~1)
assuming $\alpha=\pi/4 =\lambda_\mu$. $y^{(3\sigma)}_{b\bar b}$  is
shown with the solid line,  $y^{(3\sigma)}_{WW}$
with the
dot-dashed line, $y^{(3\sigma)}_{ZZ}$    with the dotted line,
and $y^{(3\sigma)}_{t\bar t}$    with the dashed line. (Note
$y^{3\sigma}$ is the number of years needed to accumulate a $3\sigma$
production asymmetry. Note also that the horizontal line,
$y^{3\sigma}=1$, is drawn to serve as a point of reference.)

\bigskip
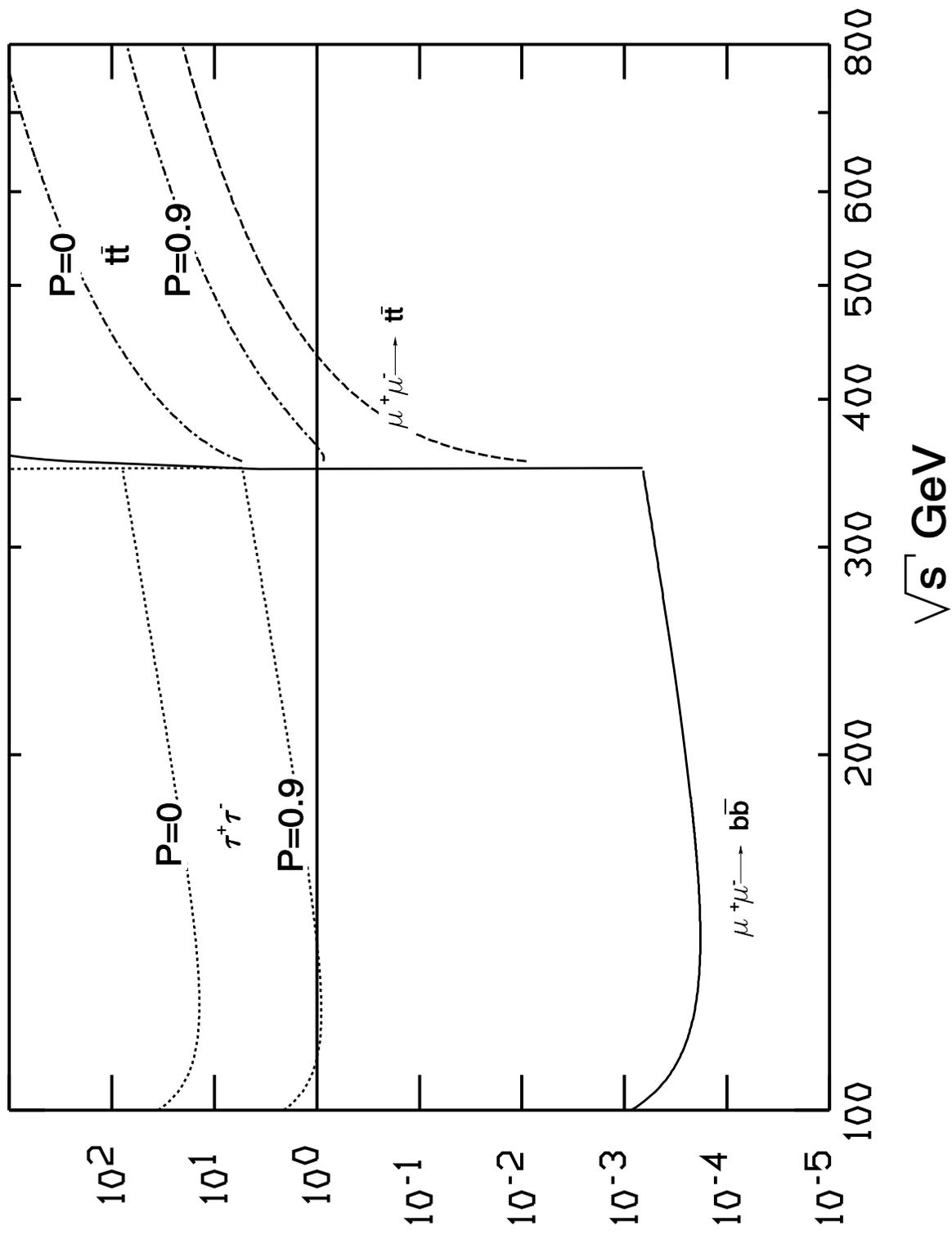
\noindent{Figure 3:}
\medskip

The value of $y^{(3\sigma)}$ for case 2 assuming $\lambda_\mu=
\lambda_\tau =\lambda_t=\pi/4$. $y^{(3\sigma)}_{b\bar b}$
is shown  with
the solid line,  and $y^{(3\sigma)}_{t\bar t}$
with the dashed
line. The results for decay asymmetries are also shown: $\hat
y^{(3\sigma)}_{t\bar t}$    with the dot-dashed line and $\hat
y^{(3\sigma)}_{\tau\tau}$    with the dotted line. In both of the
latter cases the lower curve is for $P=.9$  and the upper curve is for
$P=0$. See captions to Figures~1 and 2.

\bigskip
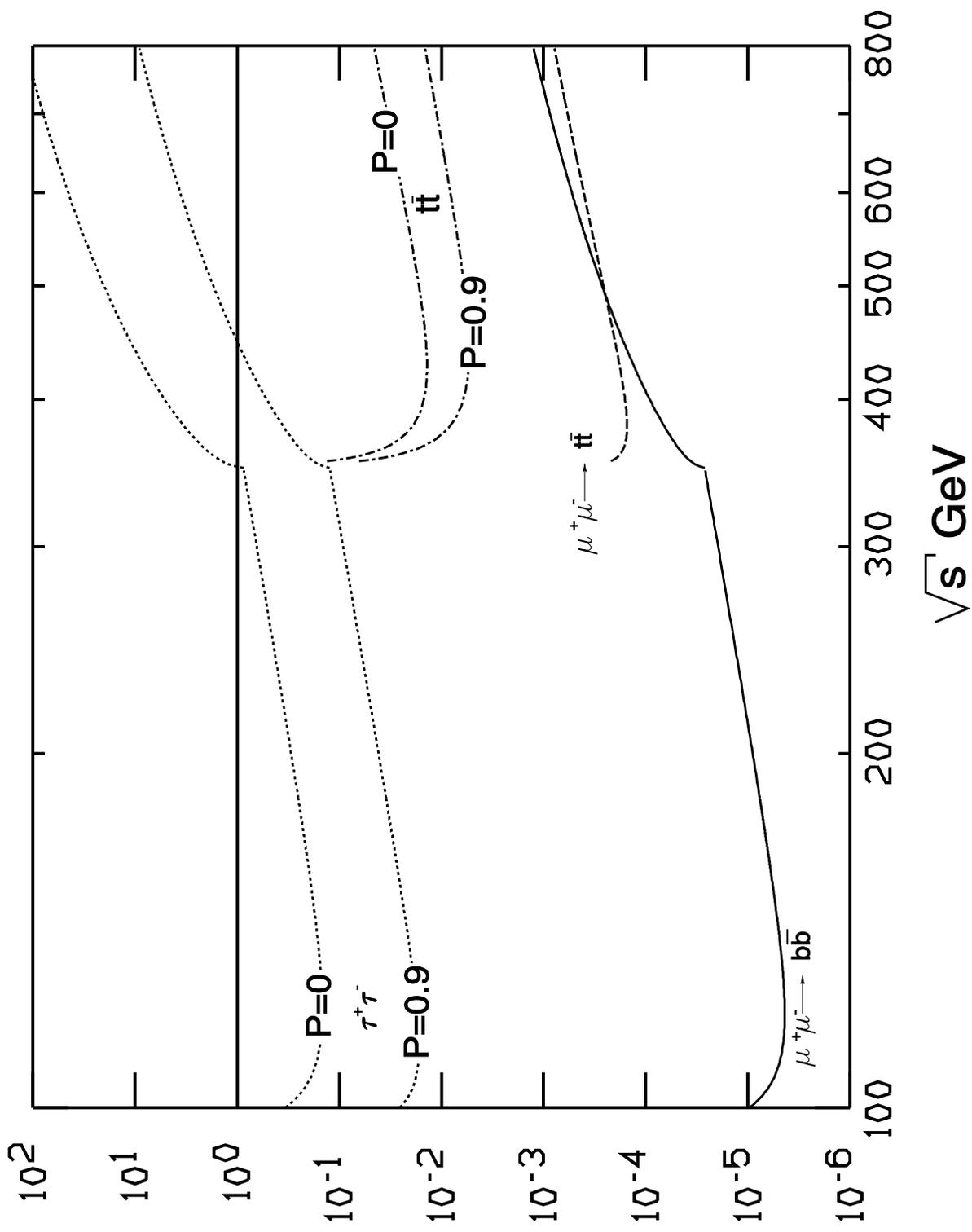
\noindent{Figure 4:}
\medskip

The value of $y^{(3\sigma)}$ for case 3 assuming $\lambda_\mu=
\lambda_\tau= \lambda_t=\pi/4$. $y^{(3\sigma)}_{b\bar b}$  is shown with
the solid line,  and $y^{(3\sigma)}_{t\bar t}$    with the dashed
line. The results for decay asymmetries are also shown: $\hat
y^{(3\sigma)}_{t\bar t}$    with the dot-dashed line and $\hat
y^{(3\sigma)}_{\tau\tau}$    with the dotted line. Again in both
of the latter cases the lower curve is for $P=.9$  and the upper curve
is for $P=0$. See captions to Figures~1 and 2.

\pagebreak


\begin{thebibliography}{99}
\bibitem{kuhnone} In principle if the dynamics of the hadronic state
are fully understood,
the spin of the $\tau$ may be fully reconstructed
in terms of the momenta of the final state mesons.
(J. H. K\"uhn, private communication).
\bibitem{simsug} A similar suggestion to use the longitudinal
polarization of the muons to enhance signal over background is also made
in Ref.~\cite{cline} and in Ref.~\cite{barger}.
\end{thebibliography}
\end{document}